\documentclass[
aps,
nofootinbib,
superscriptaddress,
tightenlines,
notitlepage,
twocolumn,
showpacs,
floatfix,
]{revtex4-1}
\usepackage{amssymb,amsmath,bm,tensor,braket}
\usepackage[colorlinks]{hyperref}

\usepackage{graphicx}
\usepackage{dcolumn}
\usepackage{bm}
\usepackage{xcolor}

\newcommand{\PlanckMass}{M_{\rm Pl}}

\newcommand{\Tabriz}{\affiliation{Faculty of Physics, University of Tabriz,
Tabriz 51666-16471, Iran}}

\newcommand{\Athens}{\affiliation{Institute for Astronomy, Astrophysics, Space 
Applications and Remote Sensing, National Observatory of Athens, Lofos Nymfon, 
11852 Athens, Greece,}}
\newcommand{\Chile}{\affiliation{Departamento de Matem\'{a}ticas, Universidad Cat\'{o}lica del Norte, Avda. Angamos 0610, Casilla 1280 Antofagasta, Chile}}
\newcommand{\China}{\affiliation{CAS Key Laboratory for Researches in Galaxies 
and Cosmology,
Department of Astronomy, University of Science and Technology of China, Hefei,
Anhui 230026, P.R. China}}

\begin{document}

\preprint{APS/123-QED}

\title{Aether-quasi-dilaton massive gravity}

\author{Sobhan Kazempour}\email{s.kazempour@tabrizu.ac.ir}\Tabriz
\author{Amin Rezaei Akbarieh}\email{am.rezaei@tabrizu.ac.ir}\Tabriz
\author{Emmanuel N. Saridakis}\email{msaridak@noa.gr}\Athens\Chile\China


\begin{abstract}
Although quasi-dilaton massive gravity is a well-defined gravitational 
theory, it exhibits
 instabilities and suffers from the strong coupling problem. In this work we 
construct an extension of the theory, through the inclusion of the aether field. 
Focusing on flat  Friedmann-Lema\^itre-Robertson-Walker  geometry, we show the 
existence of exact, self-accelerating solutions  at the 
background level, characterized by an effective
 cosmological constant arising from the graviton mass. 
 Additionally, we perform a detailed perturbation analysis,  
investigating separately the tensor, vector, and scalar 
perturbations, extracting the dispersion 
relation of gravitational waves, and determining the stability 
conditions for vector and scalar sectors.  As we show, there are always regions 
in the parameter space in which the obtained solutions are free from ghost 
instabilities, as well as from the strong coupling problem.
Hence, although the aether field does not play an 
important role in the background self-accelerating solutions, it does play a 
crucial
role in the alleviation of the perturbation-related problems of the simple 
quasi-dilaton massive gravity.

\end{abstract}


\maketitle


\section{\label{sec:intro}Introduction}

The origin of the late-time accelerated expansion of the Universe, supported by 
accumulating observational data  from supernova Ia 
\cite{Phillips:1993ng,SupernovaSearchTeam:1998fmf}, cosmic microwave background 
(CMB) radiation \cite{Planck:2015fie,WMAP:2003elm}, baryon acoustic 
oscillations \cite{Beutler:2011hx,SDSS:2009ocz}, etc,
is one of 
the  essential   issues of the standard cosmological paradigm.   It is 
noticeable that the accelerated expansion   can been explained in the context 
of general relativity, which is a unique theory of a massless Lorentz-invariant 
spin-2 particle in four dimensions \cite{Weinberg:1965rz}, by considering the 
cosmological constant \cite{Weinberg:1988cp,Peebles:2002gy}, or the 
         dark energy sector
\cite{Copeland:2006wr,Carroll:2003st,Cai:2009zp,Bamba:2012cp}. 

On the other hand, one may explain the accelerated expansion through the 
paradigm of modified gravity 
\cite{Nojiri:2010wj,Clifton:2011jh,CANTATA:2021ktz,Ishak:2018his,
Abdalla:2022yfr}. One direction withing this framework is curvature-based 
gravity,  such as   $f(R)$ gravity 
\cite{Starobinsky:1980te},   $f(G)$
gravity \cite{Nojiri:2005jg},   $f(P)$ 
gravity 
\cite{Erices:2019mkd}, Lovelock gravity \cite{Lovelock:1971yv},   
Horndeski/Galileon scalar-tensor theories 
\cite{Horndeski:1974wa,Deffayet:2009wt}  etc. Alternatively one may proceed 
with torsion-based modified gravity, such as  $f(T)$ gravity 
\cite{Bengochea:2008gz,Cai:2015emx},   $f(T,T_{G})$ gravity 
\cite{Kofinas:2014owa},   $f(T,B)$ gravity 
\cite{Bahamonde:2015zma},
  scalar-torsion theories \cite{Geng:2011aj} etc.

One interesting sub-class of  gravitational modification is massive gravity,   
in which the propagation of gravity corresponds to a spin-2 massive graviton 
\cite{deRham:2010ik,deRham:2010kj,Hinterbichler:2011tt,deRham:2014zqa,
Hassan:2011hr,Hassan:2011zd}.
The first analysis to describe the massive spin-2 field theory was performed  
by Fierz and Pauli in 1939. They presented the unique Lorentz-invariant linear 
theory without ghosts in a flat spacetime, by considering a massive spin-2 
particle that consists of a specific combination of the mass terms, resulting 
to five physical degrees of freedom \cite{Fierz:1939ix}.
In the following decades, van Dam, Veltman and Zakharov found that the 
Fierz-Pauli theory in the massless limit  does not 
reduce to the massless theory, since there is a discontinuity (van 
Dam-Veltman-Zakharov  (vDVZ) discontinuity) 
\cite{vanDam:1970vg,Zakharov:1970cc}. Hence, Vainshtein argued that in order to 
avoid the vDVZ discontinuity the theory should be extended to the nonlinear 
level \cite{Vainshtein:1972sx}.
However, Boulware and Deser reported that the nonlinear theory  of Fierz and 
Pauli exhibits a ghost, namely an instability that was later called the 
Boulware-Deser ghost \cite{Boulware:1972yco}.
Finally, de Rham, Gabadadze, and Tolley (dRGT) presented a fully nonlinear 
massive gravity without the Boulware-Deser ghost  in a certain 
decoupling limit, namely the dRGT massive gravity 
\cite{deRham:2010ik,deRham:2010kj}.

While   dRGT massive gravity   can explain the accelerated expansion of the 
Universe for an open   Friedmann-Lema\^itre-Robertson-Walker (FLRW) geometry, 
it cannot present any solutions for homogeneous and isotropic Universe 
\cite{DeFelice:2012mx}.
Furthermore, due to the strong coupling problem  and the nonlinear ghost 
instability, the scalar and vector perturbations would vanish 
\cite{Gumrukcuoglu:2011zh}. Thus, the quasi-dilaton massive gravity theory 
has been introduced in order to solve these problems  
\cite{DAmico:2012hia,Gannouji:2013rwa} (see also   
\cite{DeFelice:2013dua,Mukohyama:2014rca,Tamanini:2013xia,
EmirGumrukcuoglu:2014uog,Cai:2013lqa,Kahniashvili:2014wua,Cai:2014upa,
Gumrukcuoglu:2016hic,
Gumrukcuoglu:2017ioy,Gumrukcuoglu:2020utx,DeFelice:2021trp,Akbarieh:2021vhv,
Akbarieh:2022ovn,Aslmarand:2021qwn,Kazempour:2022let}). Nevertheless, in the 
quasi-dilaton massive gravity there is an instability in the scalar 
perturbation analysis 
\cite{Gumrukcuoglu:2013nza,Haghani:2013eya,DAmico:2013saf}.  

In order to solve the above issue, in this work 
we introduce the aether-quasi-dilaton massive gravity.  In fact, we introduce 
the new extension of the quasi-dilaton massive gravity   by considering the 
aether field in the action, and this novel extension  exhibits instability-free 
perturbations. We mention here that   although   Lorentz violation has not been 
experimentally observed \cite{Mattingly:2005re}, it cannot be theoretically 
excluded, and thus  gravitational models which violate Lorentz symmetry have 
been studied  in detail 
\cite{Carroll:2009mr,Blas:2007zz,Haghani:2015iva,Almeida:2017zzo, 
dePaulaNetto:2017fpo,Mewes:2019dhj}. In these lines,  Einstein-aether theory is 
one of the Lorentz violating   theories that has attracted   attention 
\cite{Jacobson:2004ts,Jacobson:2010mx,Carruthers:2010ii,Garfinkle:2011iw, 
Heinicke:2005bp}. In several studies, the Einstein-aether theory has been used 
to describe different aspects of the gravitational system 
\cite{Ding:2015kba,Chan:2019mdn,Kucukakca:2020ydp,Lin:2017anh,Akhoury:2016mrc}. 
We mention that this theory is a second-order one, and can explain the 
classical limit of Horava-Lifshitz gravity \cite{Horava:2009uw}.

 In the following we will show the existence of self-accelerating solutions, 
and we will perform the perturbation analysis for the aether-quasi-dilaton 
massive gravity. In particular, in the perturbations analysis we will 
extract the modified dispersion relation of gravitational waves, and we will 
 present the stability conditions of vector and scalar perturbations. The paper 
is organized as follows. In Sec. \ref{sec:1}  we present the 
aether-quasi-dilaton massive gravity, and we derive the background 
equations of motion, extracting self-accelerating solutions. In 
Sec. \ref{sec:5}  we present the cosmological perturbations analysis, which 
consist of tensor, vector, and scalar perturbations.  Finally, in Sec. 
\ref{sec:6} we summarize the obtained results. 
Throughout the manuscript  we consider natural units, where $c = 
\hslash = 1$ and $\PlanckMass^{2}\equiv 8\pi G =1$, with
$G$ the Newton's constant. 

\section{Aether-Quasi-dilaton massive gravity}\label{sec:1}

In this section  we introduce the new extension of  quasi-dilaton massive 
gravity, which is constructed by adding the action of aether field.  
The total action is:
\begin{eqnarray}\label{1}
S_{Total}= S_{QDMG}+S_{Aether}.
\end{eqnarray}
The quasi-dilaton massive gravity theory   includes 
the massive
graviton term and the quasi-dilaton term \cite{DAmico:2012hia}, namely
it has the action
\begin{eqnarray}
S_{QDMG}=\frac{1}{2}\int d^{4} x \Bigg\{\sqrt{-g}\bigg[R - \omega g^{\mu\nu}\partial_{\mu}\sigma \partial_{\nu} \sigma \nonumber\\ + 2{m}_{g}^{2}U(\mathcal K)  \bigg]\Bigg\},
\end{eqnarray}
where $R$ is the Ricci scalar, $\omega$ is a dimensionless constant, $\sigma$ is 
a scalar field,  $g_{\mu\nu}$ is the physical dynamical metric and $\sqrt{-g}$ 
is its determinant. Note that   the origin of the graviton mass $m_{g}$ 
is the potential $U$ which consists of three parts, i.e.
\begin{equation}\label{Upotential1}
U(\mathcal{K})=U_{2}+\alpha_{3}U_{3}+\alpha_{4}U_{4},
\end{equation}
with  $\alpha_3$ and $\alpha_4$   dimensionless free parameters. In the above 
expression we have \cite{deRham:2010kj}
\begin{eqnarray}\label{Upotential2}
 U_{2}&=& \frac{1}{2} \big( [\mathcal{K}]^{2}-[\mathcal{K}^{2}]\big) ,
 \nonumber\\
 U_{3}&=& \frac{1}{6} \big( [\mathcal{K}]^{3}-3[\mathcal{K}][\mathcal{K}^{2}]+2[\mathcal{K}^{3}] \big) ,
 \nonumber\\
 U_{4}&=& \frac{1}{24} \big( [\mathcal{K}]^{4}-6[\mathcal{K}]^{2}[\mathcal{K}^{2}]+8[\mathcal{K}][\mathcal{K}^{3}]+3[\mathcal{K}^{2}]^2-6[\mathcal{K}^{4}]\big) , \nonumber\\
\end{eqnarray}
where  "$[\cdot]$'' is construed as the trace of the tensor
inside the brackets. Note that the building block tensor
$\mathcal{K}$ can be defined as
\begin{equation}\label{K}
\mathcal{K}^{\mu}_{\nu} = \delta^{\mu}_{\nu} -
e^{\sigma}\sqrt{g^{\mu\alpha}f_{\alpha\nu}},
\end{equation}
where $ f_{\alpha\nu}$ is the fiducial metric    defined through
\begin{equation}\label{7}
f_{\alpha\nu}=\partial_{\alpha}\phi^{c}\partial_{\nu}\phi^{d}\eta_{cd},
\end{equation}
with   $\eta_{cd}$ the Minkowski metric ($c,d= 0,1,2,3$) and 
$\phi^{c}$   the Stueckelberg fields which are introduced to restore general 
covariance. Notice that the theory is invariant under the global dilation 
transformation  $\sigma\rightarrow\sigma+\sigma_{0}$ \cite{DAmico:2012hia}.

In addition, the aether action in   (\ref{1}) corresponds to the aether field 
$u^{\mu}$, namely \cite{Jacobson:2004ts}
\begin{eqnarray}
&&\!\!\!\!\!\!\!\!\! 
S_{Aether}=-\frac{1}{2}\int d^{4} x 
\Bigg\{\sqrt{-g}\big[\beta_{1}(\sigma)u^{\nu ;\mu}u_{\nu 
;\mu}\nonumber\\
&& \ \ \ \ \ \ \ \ \ \ \ \ \ \ \   \ \ \   
+\beta_{2}(\sigma)(g^{\mu\nu}u_{\mu ;\nu})^{2}+\beta_{3}(\sigma)u^{\nu 
;\mu}u_{\mu ;\nu}\nonumber\\
&& \ \ \ \ \ \ \ \ \ \ \ \ \ \ \ \ \ \  
+\beta_{4}(\sigma)u^{\mu}u^{\nu}u_{;\mu}u_{\nu}-\lambda(u^{\mu}u_{\nu}+1) 
\big]\Bigg\},
\end{eqnarray}
where $\beta_{1}(\sigma), \beta_{2}(\sigma), \beta_{3}(\sigma)$ and 
$\beta_{4}(\sigma)$ are  the coefficient functions that define the coupling 
between the aether field and the scalar field. 
Lastly, $\lambda$ should be considered as a Lagrange multiplier, which   
ensures the aether-field unitarity, namely $u^{\mu}u_{\mu}+1=0$ 
\cite{Kanno:2006ty,Paliathanasis:2021zry}.

\subsection{Background cosmological evolution}\label{sec:2}

Let us apply the above theory in an FLRW metric at the background level.  The 
dynamical and fiducial metrics are expressed as
\begin{align}
\label{DMetric}
g_{\mu\nu}&={\rm diag} \left[-N^{2},a^2,a^2,a^2 \right], \\
\label{FMetric} 
f_{\mu\nu}&={\rm diag} \left[-\dot{f}(t)^{2},1,1,1 \right],
\end{align}
with $a$ the scale factor and   $N$   the lapse function of the
dynamical metric, which  relates
the coordinate-time $dt$ to  the proper-time $d\tau$ via $d\tau=Ndt$
\cite{Scheel:1994yr,Christodoulakis:2013xha}. Moreover, the function $f(t)$ is 
the Stueckelberg scalar function, with $\phi^{0}=f(t)$ and
$\frac{\partial\phi^{0}}{\partial t}=\dot{f}(t)$ \cite{Arkani-Hamed:2002bjr}.

According to the above discussion, we result to the total Lagrangian 
\begin{eqnarray}
&&
\!\!\!\!\!\!\!\!\!
\mathcal{L}= \frac{-3 a\dot{a}^{2}}{N}\bigg[ 1 + 
\frac{A(\sigma)}{2}\bigg]+\frac{\omega 
a^{3}}{2N}\dot{\sigma}^{2} \nonumber\\
&&
+m_{g}^{2}\Bigg\lbrace Na^{3}(Y\!-\!1)
 \bigg[3(Y\!-\!2)\nonumber\\
&& 
\ \ \ \ \ \ \ \ \  
-(Y\!-\!4)(Y\!-\!1)\alpha_{3} -(Y\!-\!1)^{2}\alpha_{4}\bigg]
\nonumber\\
&&
+\dot{f}(t)a^{4}Y(Y\!-\!1)\bigg[3-3(Y\!-\!1)\alpha_{3}+(Y\!-\!1)^{2}\alpha_{4}
\bigg]\! \Bigg\rbrace ,
\end{eqnarray}
where
\begin{eqnarray}\label{XX}
&&A(\sigma)=\beta_{1}(\sigma)+3\beta_{2}(\sigma)+\beta_{3}(\sigma),\nonumber\\
&&
Y\equiv\frac{e^{\sigma}}{a}.
\end{eqnarray}

We proceed by considering the unitary gauge, namely $f(t)=t$, and thus  a 
constraint equation is obtained by varying with respect to $f$, i.e.
\begin{equation}\label{Cons}
\frac{\delta \mathcal{L}}{\delta f}\!= \!m_{g}^{2} \frac{d}{dt} \!\left\{  
a^{4}Y (Y\!-\!1)  [3\!-\!3(Y\!-\!1)\alpha_{3} 
\!+\!(Y\!-\!1)^{2}\alpha_{4}]\right\}=0.
\end{equation}
By varying with respect to the lapse function $N$, we obtain the Friedmann equation
\begin{eqnarray}\label{EqN}
&&\!\!\!\!\!\!\!\!\!\frac{1}{a^{3}}
\frac { \delta\mathcal{ L} }{ \delta N}= 3 H^{2}\bigg[1 + 
\frac{A(\sigma)}{2}\bigg] - \frac{\omega}{2}\big(H+\frac{\dot{Y}}{NY}\big)^{2} 
\nonumber\\ 
&& \ \ \ \ \ \ \ \ - 
m_{g}^{2}(Y-1)\bigg[(Y-4)(Y-1)\alpha_{3}\nonumber\\
&&\ \ \ \ \ \ \ \ 
\ \ \ \ \ \ \ \ \ \ \ \ \ \ \ \  
+(Y-1)^{2}\alpha_{4} -3(Y-2)
\bigg]=0.
\end{eqnarray}
Similarly, the equation of motion related to the scalar field $\sigma$ 
is given by
\begin{eqnarray}\label{EqSig}
&&\!\!\!\!\!\!\!\!\!\!\!\!\!\!\!\!\!
\frac{1}{a^{3}N}\frac{\delta\mathcal{L}}{
\delta \sigma}= - 3 H^{2} \Bigg \lbrace \frac{\omega}{N} + 
\frac{A'(\sigma)}{2} \Bigg\rbrace\nonumber\\
&& \ \ \ \ \ \
+m_{g}^{2}Y\Bigg\lbrace 
6(r+1)\big(\alpha_{4}+2\alpha_{3} 
+1\big)Y\nonumber\\
&& \ \ \ \ \ \ -(3+r)\big(3+3\alpha_{3}+\alpha_{4}\big)\nonumber\\
&& \ \ \ \ \ \ -3(3r+1)(\alpha_{4}+\alpha_{3}
)Y^{2}+4r\alpha_{4}Y^{3}\Bigg\rbrace    =0, 
\end{eqnarray}
 where $
r\equiv\frac{a}{N}$ and  $
H\equiv\frac{\dot{a}}{Na}$.
Furthermore, using the notation  of (\ref{XX}), we can write the expression
\begin{equation}
\frac{\dot{\sigma}}{N}= H+\frac{\dot{Y}}{NY}, \qquad \ddot{\sigma}=\frac{d}{dt}\Big(NH+\frac{\dot{Y}}{Y}\Big).
\end{equation}
Since the Stueckelberg field $f$ introduces a time reparametrization 
invariance, there is a Bianchi identity that relates the four equations of 
motion, namely
\begin{eqnarray}
\frac{\delta S}{\delta \sigma}\dot{\sigma}+\frac{\delta S}{\delta f}\dot{f}-N\frac{d}{dt}\frac{\delta S}{\delta N}+\dot{a}\frac{\delta S}{\delta a}=0.
\end{eqnarray}
Hence, the equation of motion corresponding to the scale factor $a$ can be 
eliminated.

\subsection{Self-accelerating background solutions}\label{sec:3}

We can now examine whether the above theory accepts    
self-accelerating solutions.
By integrating the Stueckelberg constraint  (\ref{Cons}) we obtain
\begin{eqnarray}\label{Self}
Y (Y-1)\bigg[3-3(Y-1)\alpha_{3}+(Y-1)^{2}\alpha_{4}\bigg] \propto a^{-4}. \nonumber\\
\end{eqnarray}
Hence, in an expanding universe the right-hand-side of   
(\ref{Self}) will   decrease  as  $a^{-4}$. 
Therefore, $Y$ leads to a constant value, $Y_{\rm
SA}$, which is the saturate of $Y$, and it is clear that $Y_{\rm SA}$ is a root 
of the left-hand-side of  (\ref{Self}).

As we can see, one of the solutions of (\ref{Self}) is $Y=0$. However, 
if we consider $Y=0$ the system leads to $\sigma \longrightarrow -\infty$, 
which implies that this solution leads to strong coupling in the vector and 
scalar sectors, and thus we do not study it 
\cite{DAmico:2012hia}. Hence, we have 
\begin{equation}
(Y-1)\big[3-3(Y-1)\alpha_{3}+(Y-1)^{2}\alpha_{4}\big]\bigg|_{Y=Y_{\rm SA}}=0.
\end{equation}
Moreover, another obvious solution is $Y=1$.  However, by considering this 
solution  the cosmological constant would vanish and the system would encounter 
inconsistency, and thus we do not study it either \cite{DAmico:2012hia}.
 Therefore, the two remaining solutions of   (\ref{Self}) are
\begin{equation}\label{XSa}
Y_{\rm SA}^{\pm}=\frac{3\alpha_{3}+2\alpha_{4}\pm\sqrt{9\alpha_{3}^{2}-12\alpha_{4}}}{2\alpha_{4}}.
\end{equation}
 
 The modified Friedmann equation  (\ref{EqN})  leads to
\begin{eqnarray}\label{EqFr}
3 H^{2} \bigg[ 1 + \frac{\tilde{A}}{2} -\frac{\omega}{6}\bigg] =\Lambda_{\rm SA}^{\pm}, \nonumber\\
\end{eqnarray}
where $\tilde{A}$ is the saturate of $A(\sigma)$, and $\Lambda_{\rm SA}^{\pm}$ 
is given by 
\begin{eqnarray}
&&\!\!\!\!\!\!\Lambda_{\rm SA}^{\pm}\equiv   m_{g}^{2}(Y_{\rm SA}^{\pm}-1)\Big[ 
6-3Y_{\rm SA}^{\pm}
 +(Y_{\rm SA}^{\pm}\!-\!4)(Y_{\rm SA}^{\pm}\!-\!1)\alpha_{3}\nonumber
\\&&
\ \ \ \ \ \ \ \ \ \ \ \ \ \ \ \ \ \ \ \ \ \ \ \   
+(Y_{\rm 
SA}^{\pm}\!-\!1)^{2}\alpha_{4}\Big]. 
\end{eqnarray}
Note that using  (\ref{XSa}), the above equation can be re-written  as
\begin{eqnarray}
\Lambda_{\rm SA}^{\pm}=\frac{3m^{2}_{g}}{2\alpha^{3}_{4}}\bigg[9\alpha^{4}_{3}\pm 3\alpha^{3}_{3}\sqrt{9\alpha^{2}_{3}-12\alpha_{4}}-18\alpha^{2}_{3}\alpha_{4}\nonumber\\\mp 4\alpha_{3}\alpha_{4}\sqrt{9\alpha^{2}_{3}-12\alpha_{4}}+6\alpha^{2}_{4}\bigg].
\end{eqnarray}
We solve Eq. (\ref{EqFr}) to calculate the $H^{2}$, so we have
\begin{eqnarray}
H^{2}= \frac{2 \Lambda_{SA}^{\rm \pm}}{3 \tilde{A} - \omega +6}.
\end{eqnarray}
Additionally, from  (\ref{EqSig}) we obtain  $r_{\rm SA}$ as
\begin{eqnarray}\label{rS}
r_{\rm SA}= 1 + \frac{\omega H^{2}}{m_{g}^{2}  Y_{\rm SA}^{2 \rm \pm}\big( \alpha_{3}Y_{\rm SA}^{\pm} -\alpha_{3} -2 \big)}.
\end{eqnarray}

We mention that we have acquired  a   result 
for $r_{\rm SA}$ similar with that of \cite{Gumrukcuoglu:2013nza}, which implies 
that    the aether part of the theory does not affect 
 $r_{\rm SA}$. However, we stress that there is 
not any strong coupling in this condition, and thus this theory possesses 
well-behaved self-accelerating solutions with an effective cosmological 
constant. This is one of the main results of the present work.

\begin{figure}
\centering
\includegraphics[width=7cm]{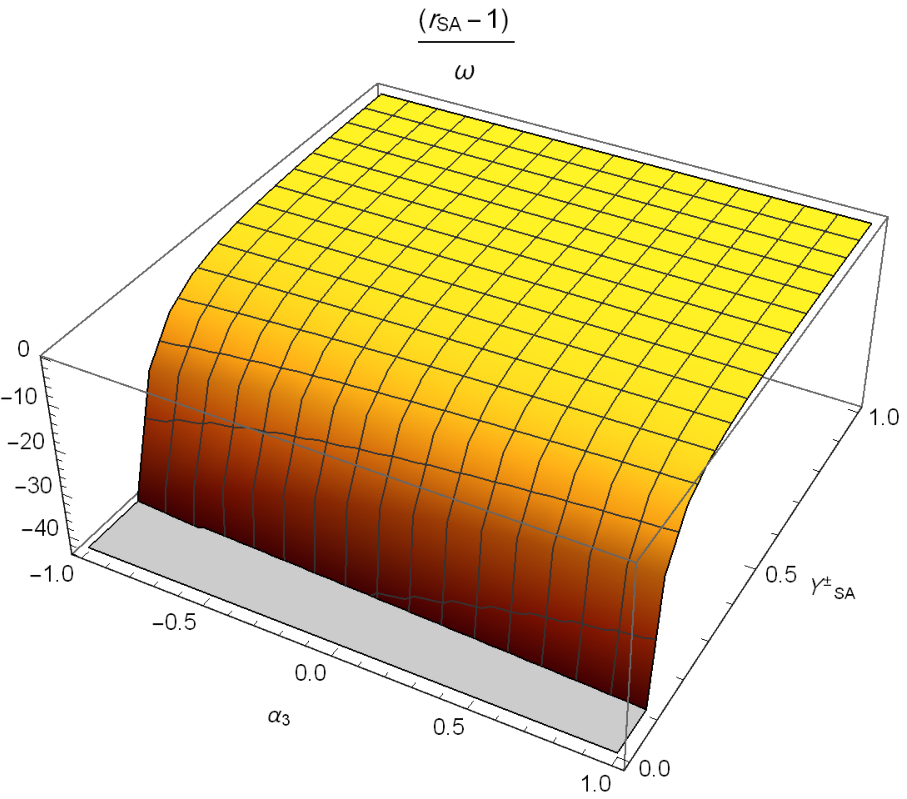}
\includegraphics[width=0.8cm]{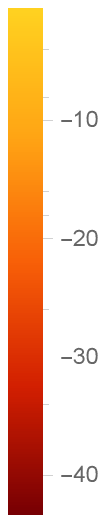}
\caption[figs]
{{\it {The quantity $\frac{r_{\rm SA}-1}{\omega}$, using (\ref{rS}), for $m_{g}/H \simeq 1$, in the case $0 < Y_{\rm SA}^{\pm} < 1$. The excluded 
regions are illustrated in grey color.}}}
\label{fig1}
\end{figure}

\begin{figure}
\centering
\includegraphics[width=7cm]{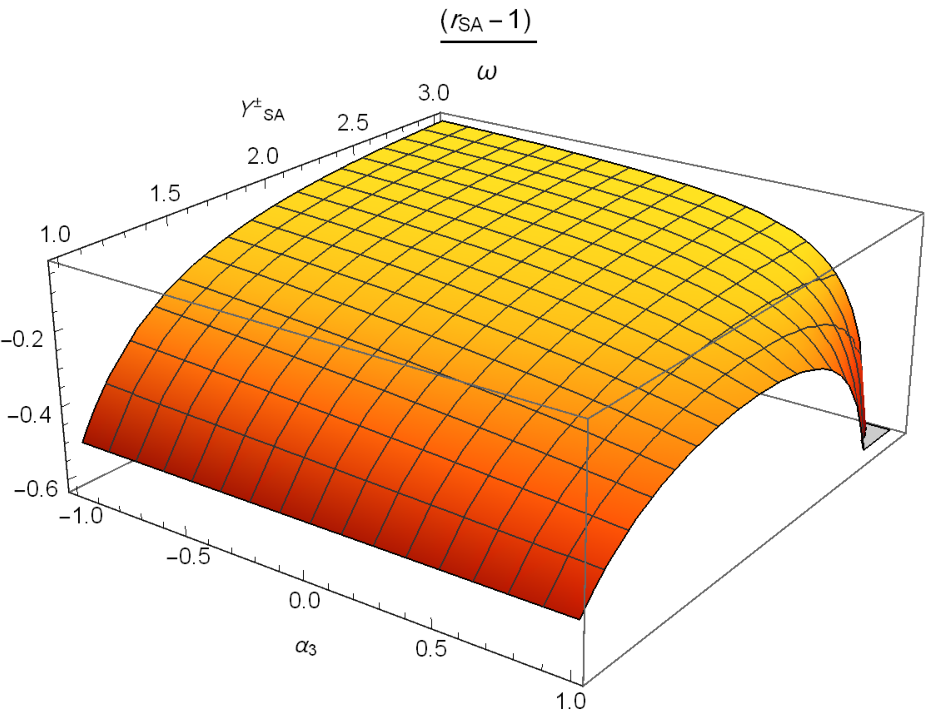}
\includegraphics[width=0.8cm]{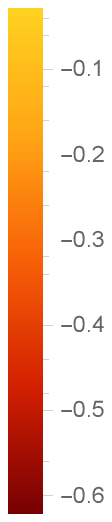}
\caption[figs]
{{\it{ The quantity $\frac{r_{\rm SA}-1}{\omega}$, using (\ref{rS}), 
for $m_{g}/H \simeq 1$, in the case $ Y_{\rm SA}^{\pm} > 1$. The excluded 
regions are illustrated in grey color.}}}
\label{fig2}
\end{figure}

In order to present the above results in a more transparent way, in
   Figs. \ref{fig1} and \ref{fig2} we illustrate the allowed parameter
regions for (\ref{rS}). Note that these figures are generated by considering 
$m_{g}/H \simeq 1$ \cite{Kahniashvili:2014wua}. We mention that by 
adjusting the value of the parameter $\alpha_{4}$  it is possible to have a 
sizeable value of   $r_{\rm SA}$.


\section{Perturbations Analysis}\label{sec:5}

In this section we perform the perturbation analysis of the scenario at hand.
The significance of such analysis    is that the stability  
conditions of the solutions can be determined, too. Since we are interested in 
quadratic perturbations, we expand the physical 
metric $g_{\mu\nu}$ in terms of small fluctuations $\delta g_{\mu\nu}$ around 
the background solution $g_{\mu\nu}^{(0)}$:
\begin{equation}
g_{\mu\nu}=g_{\mu\nu}^{(0)}+\delta g_{\mu\nu}.
\end{equation}
We keep all terms up to   quadratic order, and as usual the metric 
perturbations can be split into three parts, namely scalar, vector, and 
tensor perturbations. Thus, we have
\begin{eqnarray}
\delta g_{00}=&&-2N^{2} \Phi, \nonumber\\
\delta g_{0i}=&&Na(B_{i}+\partial_{i}B), \nonumber\\
\delta g_{ij}=&&a^{2}\bigg[h_{ij}+\frac{1}{2}(\partial_{i}E_{j}+\partial_{j}E_{i})+2\delta_{ij}\Psi \nonumber\\
&& \ \ \ \ \, +\big(\partial_{i}\partial_{j} 
-\frac{1}{3}\delta_{ij}\partial_{l}\partial^{l}\big)E\bigg].
\label{pertgener}
\end{eqnarray}
As usual, the tensor perturbations are transverse  $\partial^{i}h_{ij}=0$, 
and traceless  $h_{i}^{~ i}=0$, while the vector ones are 
transverse $\partial^{i}E_{i}=\partial^{i}B_{i}=0$. 
Notice that  all perturbations are functions of time and space, and they are 
consistent with the transformations under spatial rotations 
\cite{Kahniashvili:2014wua,DeFelice:2013tsa}.

Additionally, we consider the perturbation of the scalar field $\sigma$ as
\begin{equation}
\sigma =\sigma^{(0)}+\delta\sigma,
\end{equation}
moreover, we perturb the aether field as \cite{Zlosnik:2007bu,Battye:2017zvv},
\begin{equation}\label{upe}
u^{\mu} = u^{\mu\ (0)} + \delta u^{\mu} = \frac{1}{a}(1-\Psi, \partial^{i}V + i S_{i}),
\end{equation}
where
\begin{eqnarray}
\delta u^{\mu} = \frac{1}{a}(-\Psi, \partial^{i}V + i S_{i}),
\end{eqnarray}
here $V$ is the longitudinal scalar mode and $S_{i}$ is the transverse vector mode i.e., $\partial^{i}S_{i}=0$.
In the vector perturbations, it can be possible to restrict ourselves to the condition where the Aether field would be defined by the Khronon \cite{Blas:2010hb,Blas:2011en,Horava:2009uw,Blas:2009qj}.
The Khronometric model is a version of Einstein-Aether where the Aether field is constrained via a scalar field $\sigma$.
This way, the field can be considered as
\begin{equation}
u_{\mu} = -\frac{\partial_{\mu}\sigma}{\sqrt{-g^{\alpha\beta}\partial_{\alpha}\sigma \partial_{\beta}\sigma}},
\end{equation}
and thus the time-like unit norm constraint should be satisfied automatically. This way, the Aether is restricted to be orthogonal to a set of space-like surfaces defined by $\sigma$. At background order it can be proposed $\sigma = \sigma(t)$ and therefore from the above equation we have $u^{\mu} = (1, 0, 0,0)$. Consequently, the choice of the Khronon definition has no effect on background dynamics.
In the khronometric model, the $\sigma$ sets a preferred global time coordinate. In \cite{Blas:2010hb}, it was investigated how this model explains the low energy limit of the consistent extension of Horava gravity which is a quantum theory of gravity. At low energies, this reduces to a Lorentz-violating scalar-tensor gravity theory.\\
For the vector perturbations, so we have 
\begin{eqnarray}
\delta u_{\mu} = \frac{a}{\sigma^{'}} \bigg[ -\partial_{\mu}\delta\sigma + \partial_{\mu}\sigma \big( \Psi + \frac{\delta\sigma^{'}}{\sigma^{'}}\big) \bigg],
\end{eqnarray}
where $\delta\sigma$ is the perturbed field. The time component is then $\delta u_{0} = a \Psi$, which is a result of the time-like unit norm constraint, as in Eq. (\ref{upe}). But, if we calculate the spatial component we have
\begin{eqnarray}
\delta u_{i} = -\frac{a}{\sigma^{'}}\partial_{i}\delta\sigma \rightarrow S_{i}=0,
\end{eqnarray}
thus, there is no propagating transverse vector mode.
For the scalar perturbation of aether field, we redefine $\frac{1}{\sigma^{'}}\partial_{i}\delta\sigma = \partial_{i}V$. Thus, the scalar sector for Generalized Einstein-Aether and the Khronon should be completely equivalent \cite{Blas:2011en}.

Furthermore, as usual   the actions are expanded in Fourier plane waves, namely
$\vec{\nabla}^{2}\rightarrow -k^{2}$, $d^{3}x\rightarrow d^{3}k$, while 
 the spatial indices are raised and lowered by
$\delta^{ij}$ and $\delta_{ij}$. Lastly, since
 all calculations are performed in the unitary gauge, we do not need to specify 
gauge-invariant combinations \cite{Gumrukcuoglu:2013nza}.

\subsection{Tensor perturbations}

 We start our investigation by analyzing the tensor perturbations. Amongst 
others this  analysis  provides the speed of gravitational 
waves, and moreover it can determine  the stability of the solutions. 

For convenience, we calculate
the perturbed action at second order separately 
for the different parts. The General Relativity (GR) 
part  is written as
\begin{equation}
S^{(2)}_{\rm GR}\!=\!\frac{1}{8}\!\int \!d^{3}k   dt   a^{3}N\!\left[
\frac{\dot{h}_{ij}\dot{h}^{ij}}{N^{2}}
\!-\!\Big(\frac{k^{2}}{a^{2}}\!+\!\frac{4\dot{H}
}{N}\! +\!6H^{2} \Big)h^{ij}h_{ij}\!\right] .
\end{equation}
Additionally, the quasi-dilaton part of the perturbed action  reads as
\begin{equation}
S^{(2)}_{\rm Quasi-dilaton}=-\frac{1}{8}\int d^{3}k   dt   a^{3}N\Bigg[
\left( \frac{\omega}{N^{2}} \dot{\sigma}^{2}\right) 
h^{ij}h_{ij}\Bigg],
\end{equation}
while the aether part  
is found to be
\begin{eqnarray}
&&
\!\!\!\!\!\!\!\!\!\!\!\!\!\!\! 
S^{(2)}_{\rm Aether}=\frac{1}{16} \int  d^{3}k   dt   a^{3}N \left[
\frac{\dot{h}_{ij}\dot{h}^{ij}}{N^{2}}\right.\nonumber\\
&&\ \ \ \ \ \ \ \ \ \ \ \ \ 
\left.  - \Big(\frac{k^{2}}{a^{2}} + \frac{4\dot{H}
}{N}  + 6H^{2} \Big)h^{ij}h_{ij} \right]  A(\sigma). 
\end{eqnarray}
Finally, the massive gravity part  becomes
\begin{eqnarray}
&&\!\!\!\!\!\!\!\!
S^{(2)}_{\rm massive}= \frac{1}{8}\int d^{3}k   dt   a^{3}N 
m_{g}^{2}\bigg[(\alpha_{3}+\alpha_{4})rY^{3}\nonumber\\&&
\ \ \ \ \ \ \ \ \  \ \ \ \ \ \ \ \ \ \ \ \ \ \ \ \ \ \ \ \ \ \ 
-(1+2\alpha_{3}+\alpha_
{4})(1+3r)Y^{2}
\nonumber\\
&&\ \ \ \ \ \ \ \ \  \ \ \ \ \ \ \ \ \ \ \ \ \ \ \ \ \ \ \ \ \ \ 
+(3+3\alpha_{3}+\alpha_{4})(3+2r)Y\nonumber\\
&&\ \ \ \ \ \ \ \ \  \ \ \ \ \ \ \ \ \ \ \ \ \ \ \ \ \ \ \ \ \ \ 
-2(6+4\alpha_{3}+\alpha_{4})\bigg]h^{ij}h_{
ij}.
\end{eqnarray}

In summary, assembling the above terms, 
  the second-order    perturbed action for tensor perturbations  $S^{(2)}_{\rm 
total}=S^{(2)}_{\rm GR}+S^{(2)}_{\rm quasi-dilaton}+S^{(2)}_{\rm 
aether}+S^{(2)}_{\rm 
massive}$,  becomes
\begin{eqnarray}
&&
\!\!
\!\!\!\!\!\!\!\!\!\!
S^{(2)}_{\rm total}=\frac{1}{8}\int d^{3}k \, dt \, a^{3}N\Bigg\lbrace 
\frac{\dot{h}^{ij}\dot{h}_{ij}}{N^{2}}\bigg[1+A(\sigma)\bigg]\nonumber\\
&& \ \ \  \ \ \ \ \ \ \ \ \ \ \ \ \ \   
-\bigg\{
\frac{k^{2}} {a^{2}} [1+A(\sigma) ] +M_{\rm GW}^{2}\bigg\}
h^{ij}h_{ij}\Bigg\rbrace , 
\end{eqnarray}
where
\begin{eqnarray}\label{M_GW}
M^{2}_{\rm GW}=\left(\frac{4\dot{H}}{N}+6H^{2}\right) \left[1+A(\sigma) 
\right]+\frac{\omega}{N^{2}} \dot{\sigma}^{2}+\chi,
\end{eqnarray}
and
with
{\small{
\begin{eqnarray}
&&\chi = \frac{1}{(2Y_{\rm SA}^{\pm}-2)}\Bigg\lbrace 2 m_{g}^{2}\bigg\{ Y_{\rm 
SA}^{\pm}\bigg\{ Y_{\rm SA}^{\pm}\big[ Y_{\rm SA}^{\pm}(r_{SA}\!+\!1)\! - \!6 
\big] \!+\!6 \bigg\}  \! -\!2 
\bigg\}\nonumber\\
&&\ \ \ \ \    \
- \frac{1}{(r_{SA}-1)Y_{\rm SA}^{\pm \rm 2}N} 
\bigg\{ H^{2}\big[ Y_{\rm SA}^{\pm}(Y_{\rm SA}^{\pm} 
\!-\!3)(Y_{\rm 
SA}^{\pm}r_{SA}  \!-\!2)\!-\! 2 
\big]\nonumber\\
&& \ \ \ \ \ \ \ \ \ \ \ \ \ \ \ \ \ \ \ \ \ \ \  \ \ \ \ \ \ \ \ \ \ \ \ \ 
  \cdot \big[ 2\omega +N A'(\sigma)\big] \bigg\} \Bigg\rbrace .
\end{eqnarray}}}
The last relation is obtained using 
(\ref{XSa}) and (\ref{rS}) to substitute $\alpha_{3}$ and $\alpha_{4}$.

In summary, expression (\ref{M_GW})    determines the dispersion 
relation of 
gravitational 
waves in  aether-quasi-dilaton massive gravity.
In particular, in order to guarantee the stability  of long-wavelength 
gravitational waves, the mass square of gravitational waves  should 
be positive, namely $M_{\rm GW}^{2}>0$.

\subsection{Vector perturbations}

We proceed by performing the vector perturbation analysis.
We consider
\begin{eqnarray}\label{Bi}
B_{i}=\bigg[ A(\sigma) a +\frac{2 k^{2}a\big(r^{2}-1\big)}{2 a^{2} H^{2}\omega 
+k^{2}\big(r^{2}-1\big)} \bigg]\frac{\dot{E}_{i}}{4N}.
\end{eqnarray}
Note that the field $B_{i}$ is   non-dynamical,  and thus we handle it as an 
auxiliary field in the main action. Thus, a single propagating vector is 
obtained, namely
\begin{eqnarray}\label{AVc}
S_{\rm vector}^{(2)}=\frac{1}{8}\int d^{3}k  dt  a^{3}N 
\bigg(\frac{\beta}{N^{2}} |\dot{E}_{i}|^{2} -\frac{k^{2}}{2}M_{\rm GW}^{2}|E_{i}|^{2}\bigg),\nonumber\\
\end{eqnarray}
where
\begin{eqnarray}
\beta = \frac{k^{2}}{2}\bigg[ 1+\frac{k^{2}(r^{2}-1)}{2 a^{2}H^{2}\omega} \bigg]^{-1}.
\end{eqnarray}
It should be pointed out that in the case  $\frac{r^{2}-1}{\omega}\geq 0$, we 
have no critical momentum scale. On the other hand, for
$\frac{r^{2}-1}{\omega}<0$  we have a critical momentum scale which is 
$k_{c}^{2}=\frac{2a^{2}H^{2}\omega}{1-r^{2}}$, to avoid a ghost.
We mention that the physical critical momentum scale is vital  in order to 
acquire stability, and   this   scale should 
be above the ultraviolet cutoff scale of effective field theory, namely
\begin{eqnarray}\label{LamC}
\Lambda_{UV}^{2}\lesssim \frac{2 H^{2}\omega}{1-r^{2}}.
\end{eqnarray}
 
 Moreover, the canonically normalized fields are defined to determine other 
instabilities in the vector modes:
\begin{eqnarray}
\zeta_{i}=\frac{\beta E_{i}}{2}.
\end{eqnarray}
By substituting  into (\ref{AVc}), we have
\begin{eqnarray}
S=\frac{1}{2}\int d^{3}k \, dt \, a^{3}N \bigg(\frac{|\dot{\zeta_{i}}|^{2}}{N^{2}}-c_{V}^{2}|\zeta_{i}|^{2}\bigg).
\end{eqnarray}
Thus, the sound speed for vector modes becomes
\begin{eqnarray}\label{c_V}
c_{V}^{2}&&=M_{GW}^{2}(1+u^{2})-\frac{H^{2}u^{2}(1+4u^{2})}{(1+u^{2})^{2}},
\end{eqnarray}
where the dimensionless quantity  $u^{2}$ is
\begin{eqnarray} 
u^{2} && \equiv \frac{k^{2}(r^{2}-1)}{2 a^{2}H^{2}\omega}.
\end{eqnarray}

 Lets us proceed by elaborating  the stability conditions. 
Observing the first part of (\ref{c_V}) we deduce
that if   $M_{GW}^{2}<0$ and $u^{2}>0$ then we encounter tachyonic 
instability. In order to avoid this condition, one requires
\begin{eqnarray}
&&\Lambda_{UV}^{2} \lesssim \frac{2 H^{2}\omega}{r^{2}-1}, 
\end{eqnarray}
in the case $ \frac{(r^{2}-1)}{\omega}>0$.
It is interesting to note that if all physical momenta are considered below  the 
UV cut-off $\Lambda_{UV}$, then the rate of instability growth would be lower 
than the cosmological scale.
Furthermore, by looking at the second part of (\ref{c_V}), two cases can be 
ascertained.
Firstly, if we consider $u^{2}>0$,  there are not any instabilities faster than 
the Hubble expansion. On the other hand, for  $u^{2}<0$, due to the 
no-ghost condition   (\ref{LamC}), in order to avoid instabilities we 
require  $|u^{2}|\lesssim \frac{k^{2}}{a^{2}}\frac{1}{\Lambda_{UV}^{2}}$. 
Hence, we have no instabilities in the second part of   (\ref{c_V}).
In summary,  in order  to maintain the stability 
of the vector modes, we demand $c_{V}^{2}>0$ and $M_{GW}^{2}>0$.

\subsection{Scalar perturbations}

\begin{figure}
\centering
\includegraphics[width=7cm]{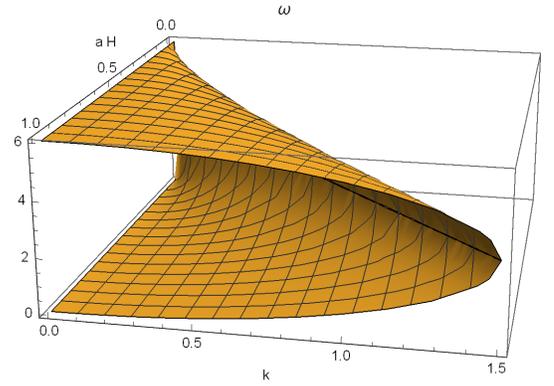}
\caption[figs]
{
{\it{ The region corresponding to absence of ghost instabilities in the scalar 
perturbations,
 corresponding to positive  determinant   (\ref{69}).}}}
\label{fig3}
\end{figure}

We proceed to the investigation of scalar perturbations, which are crucial for 
the  growth of the Universe structure. Observing the    
perturbation form (\ref{pertgener}), we 
  can handle   $\Phi$ and $B$ as auxiliary fields, since they 
are free of time derivatives. In particular, we have
\begin{eqnarray}
B=\frac{r^{2}-1}{\omega a H^{2}}\bigg[ H \big( \omega \delta\sigma -2\Phi 
\big)+\frac{1}{3N} \big( k^{2}\dot{E}+6\dot{\Psi} \big) \bigg],
\end{eqnarray}
 and
\begin{widetext}
\begin{eqnarray}
\Phi =\frac{A(\sigma)\big( k^{2}\dot{E}+6\dot{\Psi} \big)}{12 H N} \bigg[ 1- \Psi +V \bigg] + \frac{1}{48 k^{2}(r^{2}-1) - 12H^{2}a^{2}\omega (\omega-6)}\Bigg\lbrace 4 k^{4}\omega^{2}\big(E \nonumber\\+3\big) \bigg( 2 k^{2} \big( r^{2} - 1 \big) -\frac{3 a^{2}H^{2}\omega}{r-1} \bigg) \delta\sigma + 12 \omega \bigg( 2 k^{2} + \frac{3 a^{2}H^{2}\omega}{r-1} \bigg)\Psi \nonumber\\- \frac{12 H a^{2}\omega \big( \omega \delta\dot{\sigma} - 6 \dot{\Psi} \big)}{N} + \frac{8 k^{2}(r^{2}-1)\big( k^{2}\dot{E} +6 \dot{\Psi} \big)}{H N} \Bigg\rbrace . \nonumber\\
\end{eqnarray}
\end{widetext}
Hence, substituting the above expressions into the
action, we remain with four fields, namely $E$, $\Psi$, $V$ and
$\delta\sigma$. In addition, we can use another non-dynamical combination, i.e.
\begin{eqnarray}
\tilde{\Psi}= \frac{1}{\sqrt{2}}(\Psi +\delta\sigma).
\end{eqnarray}
Since $\tilde{\Psi}$ is free of time derivatives, namely
\begin{widetext}
\begin{eqnarray}
\tilde{\Psi}=&& \bigg( 1+ \frac{A(\sigma)} {12 N r}\bigg[ 144 a^{2}H^{2}N(r+1)\tilde{\delta\sigma} -k^{2}Nr ( 6\tilde{\delta\sigma} +\sqrt{2} E ) +3 a^{2}H \big( 12(r+1) \dot{\tilde{\delta\sigma}} +\sqrt{2}r \dot{E} \big) \bigg] \bigg)^{-1} \nonumber\\ &&\Bigg\lbrace - \frac{2\sqrt{2}k^{4}E}{3 \big( 4 k^{2} - a^{2} H^{2} (6 - \omega) \omega \big)} + \bigg( - k^{2} - \frac{24 a^{2} H^{2}}{r (r - 1)}  + \frac{2 a^{2}H^{2}k^{2}\big( - \omega^{2} + r (48 - (6 - \omega )\omega) \big)}{(r - 1)\big( 4 k^{2} - a^{2}H^{2}(6 - \omega )\omega \big)} \bigg) \tilde{\delta\sigma} \nonumber\\&& + \frac{2 a^{2}H}{N}\bigg( \frac{3}{r}  + \frac{(6 - \omega ) \big( 2 k^{2} (r - 1) + 3 a^{2}H^{2}\omega \big) }{(r -1)\big( 4 k^{2} - a^{2}H^{2}(6-\omega)\omega \big)} \bigg) \dot{\tilde{\delta\sigma}} + \frac{\sqrt{2}a^{2}H k^{2}(6-\omega)\dot{E}}{3N\big( 4 k^{2}- a^{2}H^{2}(6-\omega)\omega \big)} \nonumber\\&& + \frac{A(\sigma)+ V}{12 N r}\bigg( 144 a^{2} H^{2}N (r+1) \tilde{\delta\sigma} - k^{2} N r (6 \tilde{\delta\sigma} + \sqrt{2} E ) + 3 a^{2} H \big( 12 (1+r) \dot{\tilde{\delta\sigma}} + \sqrt{2}r \dot{E}\big) \bigg)  \Bigg\rbrace, \nonumber\\
\end{eqnarray}
\end{widetext}
it can be used as an auxiliary field in order to eliminate the sixth degree of freedom.  
Finally, we consider the orthogonal combination 
\begin{eqnarray}
\tilde{\delta\sigma}=\frac{1}{\sqrt{2}k^{2}}(\Psi -\delta\sigma).
\end{eqnarray}
We can now write the action in terms of $\tilde{\Psi}$,
$\tilde{\delta\sigma}$, $E$ and $V$. Introducing the notation $P \equiv
(\tilde{\delta\sigma}, E, V)$, we have
\begin{equation}
S \!= \!\frac{1}{2}\!\int\! \! d^{3}k   dt   a^{3}N\!\!\left[\!
\frac{\dot{P}^{\dagger}}{N}\mathcal{F}\frac{\dot{P}}{N}\!+\!\frac{\dot{P}^{
\dagger }} {N}\mathcal{D}P \! +\! 
P^{\dagger}\mathcal{D}^{T}\frac{\dot{P}}{N}\!-\!P^{T}\varpi^{2}P\!\right]\!.
\end{equation}
In the above expression   $\mathcal{D}$ is a real anti-symmetric $2\times 2$ matrix, 
and $\mathcal{F}$ and $\varpi^{2}$ are real symmetric $2 \times 2$ matrices. 
The components of the  $\mathcal{F}$ matrix are

\begin{eqnarray}
&&\!\!\!\!
\!\!\!\!\!\!\!\!\!\!\!\!\!\!\!\!\!\!\!\!\!\!\!\!\!\!\!\!\!
\mathcal{F}_{11}=  2 k^{4} \omega +\frac{18 k^{2}  \omega 
a^{2}H^{2}}{(r-1)^{2}} \nonumber\\
&&\!\!\!\!\!\!\!\!\!\!\!\!\!\!\!\!
- \frac{2 k^{4} a^{2}H^{2}\left[ \omega^{3} + 
(6 - \omega ) \omega^{2} r \right]^{2}}
{\left[ 4 k^{2}- (6- \omega) \omega 
a^{2}H^{2}\right] (r-1)^{2}},  
\end{eqnarray}
\begin{equation}
\mathcal{F}_{12}=  \frac{\sqrt{2}k^{4}r}{(r-1)}-\frac{2\sqrt{2}k^{6}
\left[ 
\omega^{2} + (6 - \omega )\omega r \right]}{\left[12 k^{2}\omega - 3 
(6-\omega)\omega^{2} a^{2}H^{2}\right](r-1)}  ,
\end{equation}
\begin{eqnarray}
\!\!\!\!\!\!\!\!\!\!\!\!\!\!\!\!\!\!\!\!\!\!\!\!\!\!\!\!\!
\mathcal{F}_{22}=    \frac{k^{4}\omega}{36} -\frac{k^{4} 
a^{2}H^{2}(6-\omega)^{2}\omega}{144 k^{2} - 36 a^{2}H^{2} (6 - \omega )
\omega } .
\end{eqnarray}
In order to  examine the sign of the eigenvalues, we calculate the
determinant of the kinetic matrix $\mathcal{F}$ 
as
\begin{equation}\label{69}
det \, \mathcal{F}\equiv 
\mathcal{F}_{11}\mathcal{F}_{22}-\mathcal{F}_{12}^{2}= \frac{3 
k^{6}\omega^{2}a^{4}H^{4}}{\left[\omega a^{2}H^{2}-\frac{4 k^{2}}{(6 - 
\omega)}\right](r-1)^{2}}.
\end{equation}
Hence, in order to avoid   ghost instabilities in the 
scalar sector,  we require
\begin{eqnarray}\label{60}
\frac{k}{aH} < \frac{\sqrt{6\omega - \omega^{2}}}{2}.
\end{eqnarray}
In  Fig. 
\ref{fig3} we depict the region corresponding to absence of ghost 
instabilities. As we observe, in order not to have ghosts we require  
$0<\omega <6$. Moreover,  by demanding the left-hand-side 
of  (\ref{EqFr})  to be positive,  we deduce that the aether field in the 
saturate condition $\tilde{A}$ should be positive, i.e. $\tilde{A}>0$.  
 
\section{Conclusions}\label{sec:6}

In this manuscript we    presented the aether-quasi-dilaton massive gravity. 
This theory arises from the inclusion of the aether field in the framework of 
quasi-dilaton massive gravity. After constructing the action of the theory,
we extracted the general field equations and then we applied them in a flat 
FLRW geometry.

We started our analysis at the cosmological background  level, showing the 
existence of 
exact, self-accelerating solutions, characterized by an effective
 cosmological constant arising from the graviton mass. 
 
 However, the interesting 
feature of the scenario was revealed performing a detailed perturbation 
analysis. In particular, investigating separately the tensor, vector, and 
scalar perturbations we showed that the aether-quasi-dilaton massive gravity is 
free of ghost instabilities as well as of the strong coupling problem. 

In particular, concerning the  tensor perturbations we extracted the dispersion 
relation of gravitational waves. Additionally, performing the 
 vector and scalar perturbation analysis we determined the stability 
conditions.  As we saw, there are always regions in the parameter space in 
which the obtained solutions are well-behaved at both background and 
perturbation levels. 

Hence, although the aether field does not play an 
important role in the background self-accelerating solutions, it does play a 
role in the alleviation of the perturbation-related problems of the simple 
quasi-dilaton massive gravity. This result is a good motivation for further 
investigation of the theory, and in particular of its early- and late-time 
cosmological application. Since such a study lies beyond the scope of this 
first work, it is left for a future project.

\section*{Acknowledgements}
The authors are grateful to   Nishant Agarwal for   notes 
and codes related to tensor perturbationsm and would like to thank 
  Tina Kahniashvili and  A. Emir Gumrukcuolu for   useful 
comments. E.N.S.   acknowledges participation in the COST Association 
Action CA18108 ``{\it Quantum Gravity Phenomenology in the Multimessenger 
Approach (QG-MM)}''.



\end{document}